# Backward volume vs Damon-Eshbach:
# a travelling spin wave spectroscopy comparison

U.K. Bhaskar,[1]* G. Talmelli,[2,3] F. Ciubotaru,[3] C. Adelmann,[3] and T. Devolder[1]

[1]*Centre de Nanosciences et de Nanotechnologies,*
*CNRS, Université Paris-Saclay, 91120 Palaiseau, France*
[2]*Departement Materiaalkunde, KU Leuven, 3001 Leuven, Belgium*
[3]*Imec, 3001 Leuven, Belgium*
*e-mail address: umeshkbhaskar@gmail.com*

We compare the characteristics of electrically transduced Damon-Eshbach (DESWs) and backward volume (BVSWs) configurations within the same, 30 nm thick, ferromagnetic, CoFeB waveguide. Sub-micron U-shaped antennas are used to deliver the necessary in-plane and out-of-plane RF fields. We measure the spin-wave transmission with respect to in-plane field orientation, frequency and propagation distance. Unlike DESW, BVSWs are reciprocally transduced and collected for either direction of propagation, but their ability to transport energy is lower than DESWs for two reasons. This arises first because BVSW are inductively transduced less efficiently than DESWs. Also, in the range of wavevectors (~5 rad. $\mu m^{-1}$) typically excited by our antennas, the group velocity of BVSWs stays lower than that of DESW, which leads to reduced propagation ability that impact transmission signals in an exponential manner. In contrast, the group velocity of DESWs is maximum at low fields and decreases continuously with the applied field. The essential features of the measured SW characteristics are well reciprocated by a simple, 1-D analytical model which can be used to assess the potential of each configuration.

## I. INTRODUCTION

Spin-waves (SWs)[1,2] are perturbations to the magnetic order that satisfy the condition for propagation through the material/waveguide. The typical group velocity ($V_G$) of SWs - around 1-50 km/s - is similar to that of acoustic waves[3] (AWs) but 3-4 orders of magnitude smaller than that of electromagnetic waves (EMs). The corresponding reduction in wavelength ($\lambda$) of SWs/AWs at microwave frequencies translates to a smaller footprint of RF components when compared to those based on EMs. The potential for miniaturization coupled to a high intrinsic quality factor have made acoustic resonators ubiquitous in modern RF telecommunication systems[4]. SWs, on the other hand, provide over an octave frequency tuning range by modulating the direction or magnitude of the external magnetic field[5]. Additionally, SWs can be engineered to break time-reversal symmetry and non-reciprocal wave propagation – highly desired for modern RF front end applications - can be achieved using the Dzyaloshinskii-Moriya interaction[6,7]. Finally, the technological maturity and compatibility brought about by recent advances in spintronics have made SWs ideally suited for beyond CMOS computing[8,9], and non-Boolean signal processing applications in RF[10,11,12] and logic circuits[13,14,15,16].

In a thin film ferromagnetic waveguide, with in-plane magnetization, two important SW configurations can be identified[5]: Damon-Eschbach SW (DESW) and backward volume SW (BVSW) corresponding respectively to SW propagation perpendicular and parallel to the external magnetic field[1]. DESWs are surface waves with frequencies above the ferromagnetic resonance (FMR). In contrast, BVSWs are bulk modes whose negative dispersion implies that non-zero wavevectors have frequencies below the FMR[17,18,19,20]. In addition to this inherent uniqueness, BVSWs possess certain attributes which point towards a favorable role in future logic devices for example: 1) The collinearity between wave-vector and magnetization orientation of BVSWs allows excitation of SWs in the magnetic state favored by the shape anisotropy in magnetic conduits, potentially circumventing the need for an external magnetic field; 2) RF excitation with microstrip antennas results in reciprocal amplitude for both directions of BVSW propagation[21]; 3) Parametric pumping[17,22], and interaction with spin transfer torque currents[23,24] provide magnetic field-free route to modulating BVSWs. Despite these advantages, several technical challenges related to the small $V_G$ and weak excitation efficiency have limited the study of BVSWs propagation in ultra-thin film waveguides[18,23]. On the other hand, the characteristics of surface-confined DE modes have been demonstrated to scale favorably to ultra-thin, technologically relevant, CoFeB[25], and permalloy[26,20] magnetic films.

In this work, we leverage recent improvements in Gilbert damping of the CoFeB stack[25] (Fig 1(a)) and aggressively scale the dimensions of the microwave antennas (Fig 1(b)) to boost their transduction efficiency and extend it to large wavevectors[27], allowing comparison of the propagation characteristics of DESWs and BVSWs in the same thin film waveguide. The starting material for the device fabrication is a standard resistivity silicon substrate with 300 nm thick thermal $SiO_2$ deposited at its surface. A multilayer stack of Ta (3 nm)/CoFeB (30 nm)/Ta (3 nm) is sequentially sputtered to form the seed layer, ferromagnetic material, and cap layer respectively. Subsequently, a 60 nm thick SiO$_2$ layer is deposited to serve both as dielectric isolation and as hard mask for patterning 5 µm wide conduits using ion-beam etching. After definition of SW waveguides, a 300-nm-thick spin-on-carbon (SOC) layer is spin coated and recessed to act as planarization layer. A 30 nm thick SiO$_2$ layer is deposited on top of the planarized SOC layer to encapsulate the waveguides and act as dielectric isolation. Finally, Ti/Au (10 nm/100 nm) electrodes are deposited on the SiO$_2$ surface and patterned by lift-off to create multiple tap-outs, at 2.25 µm, 4.5 µm and 6.75 µm, for spin wave detection/excitation with microwave antennas.

## II. 1-D MODEL FOR SW PROPAGATION

Single wire antennas have been conventionally used for the study of spin-waves[26]. The inductive fields generated by single wire antenna decay inversely with the distance, and hence, could lead to a strong inductive coupling between the transmitting and receiving antenna. In contrast, the fields generated in a U-based antenna decay as 1/r$^2$ which leads consequently to reduced parasitic coupling. Additionally, U-shaped antennas can be designed to ensure that the transduction efficiency is maximal around any chosen non-zero wavevector ($k$). Assuming $T_{SiN} <<$ the antenna dimensions, the distribution of wavevectors accessed by the antenna can be well approximated as[26]:

$$Ant_{field}(k) \propto sin\left[\frac{1}{2}k(g+w)\right] \times sinc[\frac{k \times w}{2}]. \tag{1}$$

For the fabricated antenna geometry: width ($w = 250\ nm$) and gap-spacing ($g = 250\ nm$), the transduction is maximum around $k \sim 6\ rad.\ \mu m^{-1}$ (Supplementary Fig 1). The microwave antenna generates a strong alternating in-plane magnetic field component $H_y$ directed along the length of the spin wave conduit. There is also an RF magnetic field component $H_z$ of comparable amplitude, but it is symmetrically distributed around the electrode. For in-plane magnetized systems, the susceptibilities linking the dynamic in-plane magnetization to the out-of-plane fields ($\chi_z$), or linking the in-plane magnetization to a transverse in-plane field ($\chi_y$) are very different in amplitudes because of the precession ellipticity[25,28,29]; they differ by a factor (H+Ms)/H at the ferromagnetic resonance, where H is the internal in-plane field. Since the latter is typically small compared to the magnetization, the ratio of the susceptibilities typically exceeds a factor of ~ 5, resulting in a weaker effective excitation by out-of-plane RF fields. Note that only the microwave field components orthogonal to the static magnetization can contribute to precession of the magnetic moment and spin wave excitation. Thus, only the weak $\chi_z \times H_z$ contributes to the excitation of BVSWs. On the other hand, both $\chi_y \times H_y$ and $\chi_z \times H_z$ contribute to DESW excitation. The sign of the spin-wave wavevector $k$ determines whether these two excitations collaborate or compete with each other, resulting in non-reciprocal emission of DESW for propagation in the +$x$ and -$x$ axis, as noticed for instance in ref[5].

While the antenna dimensions determine the wavevector with maximum transduction efficiency, the mapping from wavevector to angular frequency ($\omega$) is determined by the dispersion relationship. The anisotropy of SW dispersion necessitates the definition of the wavevector with respect to the orientation of the magnetic field. In our convention, we define $k_x$ as the component of the wavevector parallel, and $k_y$ as the component of wavevector transverse to the magnetic field orientation, respectively. In a SW waveguide, the wavevector along the length of the waveguide ($k_y$ for DESWs/ $k_x$ for BVSWs) can take on continuous values, but the wavevector along the width direction ($k_x$ for DESWs/ $k_y$ for BVSWs) can take only discrete values, corresponding to confined standing wave modes ($k_x$ or $k_y = \frac{n\pi}{w}$, with $n \geq 1$). The axial symmetry of the stripe ensures that only the fields from the odd modes (n=1, 3, 5,...) can excite and detect spin waves. In our calculations, we calculate the dispersion for only the first confined mode ($k_x$ or $k_y = \frac{\pi}{w}$), neglecting the finite contribution from the other higher order odd modes. The explicit relation between frequency and wavevector ($k^2 = k_x{}^2 + k_y{}^2$) in a SW waveguide is written as [1,28,30,31]:

$$\omega = \gamma_0 \sqrt{\left[H + M_s + \frac{2A}{\mu_0 M_s}k^2 - M_s\left(1 - \frac{1-e^{-kt}}{kt}\right)\right] \times \left[H + \frac{2A}{\mu_0 M_s}k^2 + M_s\left(1 - \frac{1-e^{-kt}}{kt}\right)\left(\frac{k_y^2}{k^2}\right)\right]}, \tag{2}$$

where H incorporates the applied field and the contribution from the shape anisotropy. Writing H$_{sat} > 0$, the field needed for hard axis saturation, we have H = H$_{app}$ + H$_{sat}$ for BV and H$_{app}$ - H$_{sat}$ for DE, $\gamma_0 = \gamma \times \mu_0$ is the gyromagnetic ratio ($\gamma$=2$\pi \times$29.16 GHz/T), $t$ is the thickness (30 $nm$), $A$ is the exchange stiffness (18.6 $pJ/m$) and M$_s$ is the saturation magnetization of the CoFeB film (1.36 $MA/m$). The above equation at $k_x = 0$ and $k_y = 0$ reduces, respectively, to the DE and BV modes of a continuous film. BVSWs displays a negative $V_G$ for small $k_x$ and positive $V_G$ at larger $k_x$. The crossover is determined by the interplay between magnetic field, exchange interactions, and dipolar contributions. On the other hand, the expression for $\omega_{DESW}$ can be simplified further by neglecting the exchange interaction and considering a first order expansion for the dipolar contribution[26].

The dispersion (solid line) and group velocity (dashed line), up to experimental realizable wavevectors (10 $rad.\ \mu m^{-1}$, see Fig. S1), is plotted for the first order DESW and BVSW mode in Figure 1(c) and (d) respectively. In the case of DESWs, the frequency increases with wavevector, resulting in a positive $V_G$. The gradient of the $\omega_{DESW}$ at low fields and small $k_y$ is large, implying a large $V_G$ (up to ~15 km/s). The spread in $\omega_{DESW}$ becomes progressively narrower at larger magnetic fields and wavevectors. On the other hand, $\omega_{BVSW}$ decreases with $k_x$ and the slope of $\frac{\partial \omega_{BVSW}}{\partial k}$ is weaker implying a smaller, negative $V_G$. Around $k_x = 5\ rad.\mu m^{-1}$, corresponding to the maximum efficiency region of the antenna, the $V_G$ span from typically 3 to 6 km/s for DESW, while that of the BVSW are in the 0.4-0.8 km/s. Around $k_x = 5\ rad.\mu m^{-1}$, the group velocity decreases with the magnetic field for DESW and increases for BVSW.

Spin waves are efficiently transduced and propagated through the waveguide when the distribution of wavevectors generated by the RF currents are permitted by the dispersion relationship. SW propagation is studied in a two-port configuration, where SWs launched from one antenna travel a finite distance within the conduit before being detected at the second antenna. The reflection data at the individual ports ($S_{11}$ or $S_{22}$) provides an estimate of the spin-wave transduction efficiency and the FMR frequency. The transmission signal ($S_{21}$ or $S_{12}$) includes this information and supplements it with data about the propagation characteristics through the SW bus. The amplitude of the SW in the waveguide is exponentially attenuated in the waveguide depending on the time spent in the waveguide, the Gilbert damping ($\alpha$~$5 \times 10^{-3}$), and $M_s$ of the magnetic film. Analytically, the amplitude of SWs at a distance $d$ from generation, can be estimated from the product of the exponential SW attenuation and phase dependent oscillation[26,32]:

$$SW_{amp}(\omega) \propto \cos(-k(\omega) \times d) \times \exp\left(-d \times \gamma_0 \times \alpha \times \frac{M_s + 2H}{2V_G}\right). \quad (3)$$

The complete frequency response is calculated by multiplying the amplitude at each frequency (Eq.3) by the square of the antenna transmission function (Eq.1) and the magnetic susceptibility at resonance. The spin precession induced by propagating SWs is detected as an oscillation of the magnetic flux experienced by the microwave antenna. Hence, measuring the frequency response at a fixed magnetic field captures the SW propagation characteristics, while sweeping through different magnetic fields provides information about the SW dispersion and the frequency dependence of the susceptibility. Thus, frequency-field (FF) maps offer a complete description of the SWs. The transmission signal of BVSW was found experimentally to be small and comparable to the electromagnetic crosstalk between the two antennas. To better reveal the spin wave signal in the BVSW configuration, the magnetic field derivative of the measured frequency response is used to retain only the SW signal[26]. A low RF power of -10 dBm was used for the measurements.

## III. MEASUREMENT RESULTS AND ANALYSIS

### 1) Damon-Eschbach spin waves

A typical FF map of DESW for a propagation length of 6.75 µm is shown in Figure 2(a). Clear amplitude non-reciprocity, arising from direction dependent excitation efficiency, is observed while comparing intensity of DESW modes for positive and negative fields or equivalently when comparing the forward and backward transmission coefficients (not shown). For both directions of propagation, the phase rotations in the frequency domain shift to higher frequencies on increasing the applied field. In Fig 2(b), line-cuts of the FF map is vertically offset to distinguish the measured SW response at different fields (solid lines). The estimated $S_{21}$ from the product of Eq. 1 and Eq. 3 is also plotted (dashed line), suggesting good agreement between theory and experiment. The analytical estimate accurately captures the experimental measurement for $k < 5\ rad.\ \mu m^{-1}$. However, at large $k$, some dephasing is observed between the model and measurement, suggesting a slight discrepancy in the modelled $V_G$ at the largest k values. The separation between the maxima and minima of the SW transmission coefficient in the frequency domain is inversely correlated with the group velocity of SWs in the time domain. Quantitatively, the frequency separation ($\Delta f_p$) between two consecutive maxima corresponds to a phase rotation of $2\pi$; thus, $V_G$ is simply the product of $\Delta f_p \times d$[20]. The $V_G$ values quoted in Fig 2(b) correspond to the median value, while the error bars correspond to the standard deviation, of $V_G$ for $k$ in the range of 4-6 $rad.\ \mu m^{-1}$. The measured values are indeed slightly smaller than the expectations of Fig. 1(b). It is observed that increasing the magnetic field, progressively narrows the spread of the DESW bandwidth in the frequency domain and consecutively reduces the $V_G$ of SWs in the time domain. Assuming a damping coefficient of 0.005 - measured on the films prior to processing - the measured group velocity of 5 km/s predicts a spin-wave attenuation length of 8 µm. Thus, the DESW are expected to be detectable over all our investigated distances without any significant change in transmission amplitudes. This is indeed confirmed in the FF map of DESWs for 2.25 µm and 4.5 µm propagation distance that are shown in Fig S2(a) and S2(b).

The number of phase rotations observed is clearly proportional to the propagation distance. Additionally, by comparing the magnitude of the SW signal for different propagation distances, we can extract the loss encountered by the SW within the conduit[20]. The Gilbert damping coefficient extracted from the measured DESW attenuation length $1/(\gamma_0 \times \alpha \times \frac{M_s + 2H}{2V_G})$ was consistent with the value (0.005) obtained from ferromagnetic resonance measured on un-patterned thin films, suggesting that no deterioration was induced from the device fabrication process.

### 2) Comparison with backward volume spin waves

The FF map and line cuts of BVSW frequency response, measured for propagation distance of 2.25 µm, are shown in Fig 3 (a) and (b), respectively. The corresponding data for 4.5 µm and 6.75 µm propagation distance is shown in Fig S3(a) and (b). The experimental transmission coefficients are compared to Eq. 2 in Fig 3(b) using same material parameters as in the DE configuration. The agreement for the BVSW configuration is less satisfactory as in the DESW case, but the essential features of the spectra are reproduced. In particular, there are four noticeable differences between the BVSW and the DESW spectra: (i) the reciprocal/non-reciprocal character, (ii) the strong/weak oscillatory character of the transmission spectra, (iii) the quantitative amplitude of the spin-wave signals, and (iv) the weakness of BVSW transmission at low field/low frequency. Let us detail these points one by one.

(i) The first striking difference is indeed the reciprocal character of the transmission signals in the BVSW configuration: apart from random noise, we could confirm that the backward and forward transmission signals are indiscernible for both positive and negative fields, for all values of propagation distance.

(ii) The second striking difference between the two configurations is the rate at which the phase oscillates in the frequency domain and with respect to the propagation length. This can be explained from the anticipated contrast of group velocity (Fig. 1) at the most relevant wavevectors ($4-6\ rad.\mu m^{-1}$) between the rather slow BVSW and the faster DESW. The group velocities extracted from the experimental spectra using successive maxima of the transmission coefficient (Fig. 3b) are in line with expectations (Fig. 1c). The increase of $V_G$ is also manifest as a broadening of the frequency interval in which spin-wave are observed [this interval has a width of circa $\frac{V_G}{w+g}$ ] with magnetic field in the FF map (Fig 3a).

(iii) In addition to the contrasting group velocities attested by the much more numerous phase rotations, the striking difference is the much weaker signal (hence weaker signal to noise ratio) of the BVSW compared to the DESW configuration. When measurable, the amplitude of BVSWs signal is ~15 (best case)-50 (sensitivity limit) times weaker than that of DESWs. This amplitude difference arises partly from a weaker excitation efficiency related to the direction of the pumping RF fields and the corresponding susceptibility terms, as already discussed. However, we believe that the most substantial part of this amplitude difference arises from a more severe attenuation of the BVSW upon propagation. Let's examine the spin wave attenuation lengths in the BVSW configuration. Unfortunately, the weak BVSW signal impedes a reliable extraction of the experimental attenuation length, such that we have to partly rely on theory. With the damping value of 0.005 and the measured group velocity of 0.8 km/s at the largest fields (250 mT), the spin-wave attenuation length is anticipated to be 1.3 µm. Thus, it is no surprise that the BVSW signals degrade substantially for the large propagation distances (see Fig. S3) and that the BVSW signal quickly gets much smaller than that of the DESW configuration.

(iv) The last discerning feature is the difficulty to observe a BVSW signal at small magnetic fields/low frequencies compared to at high magnetic fields/high frequencies. Let us first exclude three possible reasons that one could invoke. First, this very weak signal at low field is not due to the antenna efficiency function (Eq. 1), because the latter has no field dependence. Second, this low transmission at low field does also not result from the susceptibility terms, as the $\chi_z$ susceptibility at resonance increases when decreasing the FMR frequency. Third, this dramatic decrease of the BVSW signal at low field/low frequency can also not be understood from any field dependence of the attenuation length, because the attenuation lengths are in fact longer at low fields for BVSW in the relevant wavevector range. For instance, for wavevectors in the range of 4-6 $rad.\ \mu m^{-1}$, the $V_G$ of BVSWs at 250 mT is 1.3 µm and it increases to 2.8 µm when the field is lower to 50 mT, such that this sole argument should render the low field/low frequency signal easier to measure than the high field signal, which is opposite to experimental findings.

We believe that the likely reason for the loss of the low field BVSW signal can not be accounted for by the crude model of Eq. 2 when the group velocity is small, as occurring for BVSW at low fields. To illustrate this point, let us discuss what would happen in a hypothetical extreme situation where the group velocity would vanish. In that hypothetical case, an emitter antenna operated at the ferromagnetic resonance frequency would emit simultaneously a wave-packet of spin-waves with all the possible wavevectors allowed by antenna, i.e. from 0 up to typically $2\pi \frac{1}{w+g}$ . Since these spin-waves would have the same frequency but different wavevectors, they would arrive at the receiving antenna with different phases spanning from 0 to $2\pi \frac{r}{w+g}$. As soon as the propagation distance $r$ gets much larger than the antenna extension $w+g$, these different spin-waves would interfere out, yielding a very low transmission signal after summation of all contributions. This hypothetical extreme situation bears some similarity with the BVSW configuration at low fields when the group velocity is sufficiently low that $\frac{V_G}{w+g}$ gets smaller that the FMR linewidth $\alpha \times \gamma_0(M_s + 2H)$, but the quantitative modeling of this effect exceeds the scope of the present study.

# IV. CONCLUSIONS

In summary, we have reported on the first electrical study on the excitation and propagation characteristics of both BVSW and DESW in the same ferromagnetic bus, and we have described their main properties using a simple, physically intuitive model. Thanks to a higher group velocity, DESWs have the clear advantage of a relatively stronger transduction efficiency, as well as longer propagation distances. BVSWs are damped at substantially faster spatial rates but display reciprocal transmission capability with potential implications for device reconfigurability and ease of operation. To circumvent the insufficient character of the excitation efficiency, inductive antennas should be replaced by spin-orbit torque antennas[25] or magnetoelectric cell-based techniques[33]. Thus, the transduction efficiency of DESW/BVSW with these techniques would determine final technological application. Fundamentally, BVSWs, by virtue of propagation in the orientation favored by the shape anisotropy of the spin-wave conduit, offer several niche benefits which scale favorably to the nanoscale. In contrast to the micron-scale spin-wave conduits used in the present study, one could harness nanoscale conduit widths in which the shape anisotropy field can be expected to be strong enough to reach both high operation frequency and reasonable BVSW group velocities in the absence of external fields, offering improved spin wave propagation capability. Using our simple one-dimensional model that accounts the essential contrasting properties of BVSWs and DESWs, their specificities could be used to overcome the drawbacks of each configuration and to design complementary SW logic circuits.

# V. SUPPLEMENTARY INFORMATION

See supplementary material for the plot of the antenna efficiency function and for the measurement of BVSW and DESW characteristics for other propagation lengths.

# VI. ACKNOWLEDGEMENTS

This work was funded through the FETOPEN-01-2016-2017 - FET-Open research and innovation actions (CHIRON project: Grant agreement ID: 801055).

[1] B. Hillebrands and A. Thiaville, *Spin Dynamics in Confined Magnetic Structures.* (Springer Verlag, 2014).

[2] A. V. Chumak, V.I. Vasyuchka, A.A. Serga, and B. Hillebrands, Nat. Phys. **11**, 453 (2015).

[3] B.A. AULD, in (ACADEMIC PRESS, INC., n.d.).

[4] R. Aigner, 2008 IEEE Ultrason. Symp. 582 (2008).

[5] R.W. Damon and J.R. Eshbach, J. Appl. Phys. **31**, S104 (1960).

[6] A. Giordano, R. Verba, R. Zivieri, A. Laudani, V. Puliafito, G. Gubbiotti, R. Tomasello, G. Siracusano, B. Azzerboni, M. Carpentieri, A. Slavin, and G. Finocchio, Sci. Rep. **6**, 1 (2016).

[7] R. Verba, I. Lisenkov, I. Krivorotov, V. Tiberkevich, and A. Slavin, Phys. Rev. Appl. **9**, 64014 (2018).

[8] S. Manipatruni, D.E. Nikonov, C.C. Lin, T.A. Gosavi, H. Liu, B. Prasad, Y.L. Huang, E. Bonturim, R. Ramesh, and I.A. Young, Nature **565**, 35 (2019).

[9] S. Manipatruni, D.E. Nikonov, and I.A. Young, Nat. Phys. **14**, 338 (2018).

[10] W.S. Ishak, Proc. IEEE **76**, 171 (1988).

[11] J.H. Collins, J.M. Owens, and C. V. Smith, Ultrason Symp Proc Phoenix Ariz 541 (1977).

[12] F. Heussner, M. Nabinger, T. Fischer, T. Brächer, A.A. Serga, B. Hillebrands, and P. Pirro, Phys. Status Solidi - Rapid Res. Lett. **12**, 1 (2018).

[13] F. Ciubotaru, G. Talmelli, T. Devolder, O. Zografos, M. Heyns, C. Adelmann, and I.P. Radu, in *2018 IEEE Int. Electron Devices Meet.* (IEEE, 2018), pp. 36.1.1-36.1.4.

[14] A. V. Chumak, A.A. Serga, and B. Hillebrands, Nat. Commun. **5**, 1 (2014).

[15] A. Khitun and K.L. Wang, J. Appl. Phys. **110**, 034306 (2011).

[16] T. Brächer and P. Pirro, J. Appl. Phys. **124**, (2018).

[17] A.A. Serga, A. V Chumak, and B. Hillebrands, J. Phys. D. Appl. Phys. **43**, 264002 (2010).

[18] N. Sato, N. Ishida, T. Kawakami, K. Sekiguchi, N. Sato, N. Ishida, T. Kawakami, and K. Sekiguchi, Appl. Phys. Lett. **104**, 032411 (2017).

[19] H.J.J. Liu, G.A. Riley, and K.S. Buchanan, IEEE Magn. Lett. **6**, 7 (2015).

[20] V. Vlaminck and M. Bailleul, Phys. Rev. B - Condens. Matter Mater. Phys. **81**, (2010).

[21] T. Schneider, A.A. Serga, T. Neumann, B. Hillebrands, and M.P. Kostylev, Phys. Rev. B **77**, 214411 (2008).

[22] T. Brächer, F. Heussner, T. Meyer, T. Fischer, M. Geilen, B. Heinz, B. Lägel, B. Hillebrands, and P. Pirro, J. Magn. Magn. Mater. **450**, 60 (2018).

[23] N. Sato, S.W. Lee, K.J. Lee, and K. Sekiguchi, J. Phys. D. Appl. Phys. **50**, (2017).

[24] V. Vlaminck and M. Bailleul, Nature **322**, 410 (2008).

[25] G. Talmelli, F. Ciubotaru, K. Garello, X. Sun, M. Heyns, I.P. Radu, C. Adelmann, and T. Devolder, Phys. Rev. Appl. **10**, 044060 (2018).

[26] F. Ciubotaru, T. Devolder, M. Manfrini, C. Adelmann, and I.P. Radu, Appl. Phys. Lett. **109**, 012403 (2016).

[27] T. Brächer, M. Fabre, T. Meyer, T. Fischer, S. Auffret, O. Boulle, U. Ebels, P. Pirro, and G. Gaudin, Nano Lett. **17**, 7234 (2017).

[28] V.E. Demidov, M.P. Kostylev, K. Rott, P. Krzysteczko, G. Reiss, and S.O. Demokritov, Appl. Phys. Lett. **95**, 10 (2009).

[29] H.J.J. Liu, G.A. Riley, and K.S. Buchanan, IEEE Magn. Lett. **6**, 7 (2015).

[30] V.E. Demidov and S.O. Demokritov, IEEE Trans. Magn. **51**, 1 (2015).

[31] M. Belmeguenai, J.P. Adam, Y. Roussigné, S. Eimer, T. Devolder, J. Von Kim, S.M. Cherif, A. Stashkevich, and A. Thiaville, Phys. Rev. B - Condens. Matter Mater. Phys. **91**, 1 (2015).

[32] O. Gladii, D. Halley, Y. Henry, and M. Bailleul, Phys. Rev. B **96**, 174420 (2017).

[33] S. Cherepov, P. Khalili Amiri, J.G. Alzate, K. Wong, M. Lewis, P. Upadhyaya, J. Nath, M. Bao, A. Bur, T. Wu, G.P. Carman, A. Khitun, and K.L. Wang, Appl. Phys. Lett. **104**, 1 (2014).

**Figures**

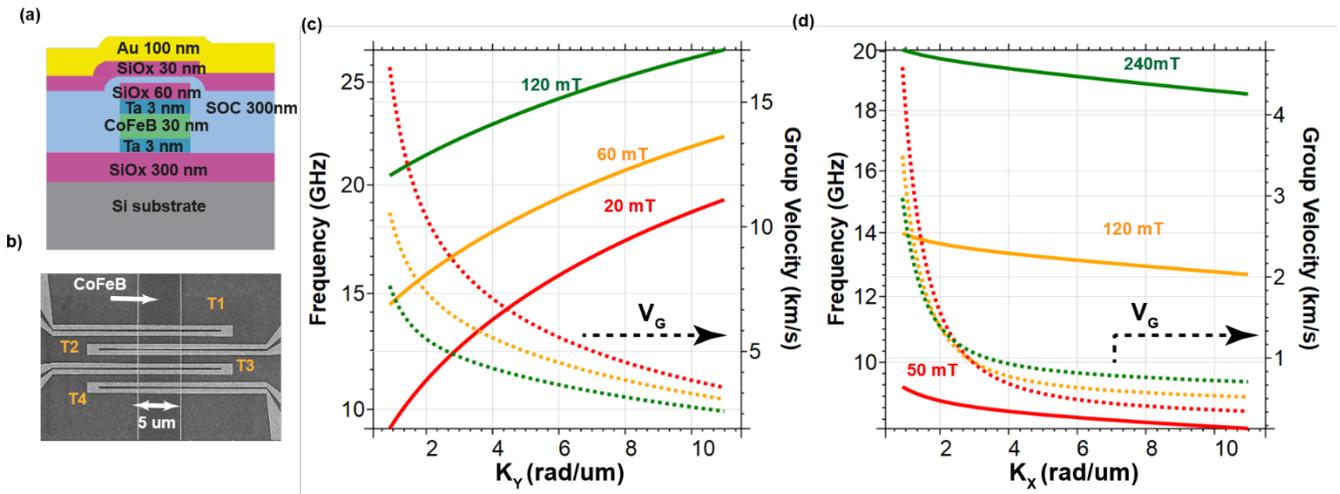

**Figure 1(a):** Complete ferromagnetic stack along with the metal antenna, isolation and buffer layers; **(b)** In-line design of spin-wave bus with multiple tap-outs using four U-shaped microwave antennas; Frequency (bold lines, left axes) and groupvelocity (dotted lines, right axes) as a function of wavevector for the first-order **(c)** DESWs and **(d)** BVSWs, respectively.

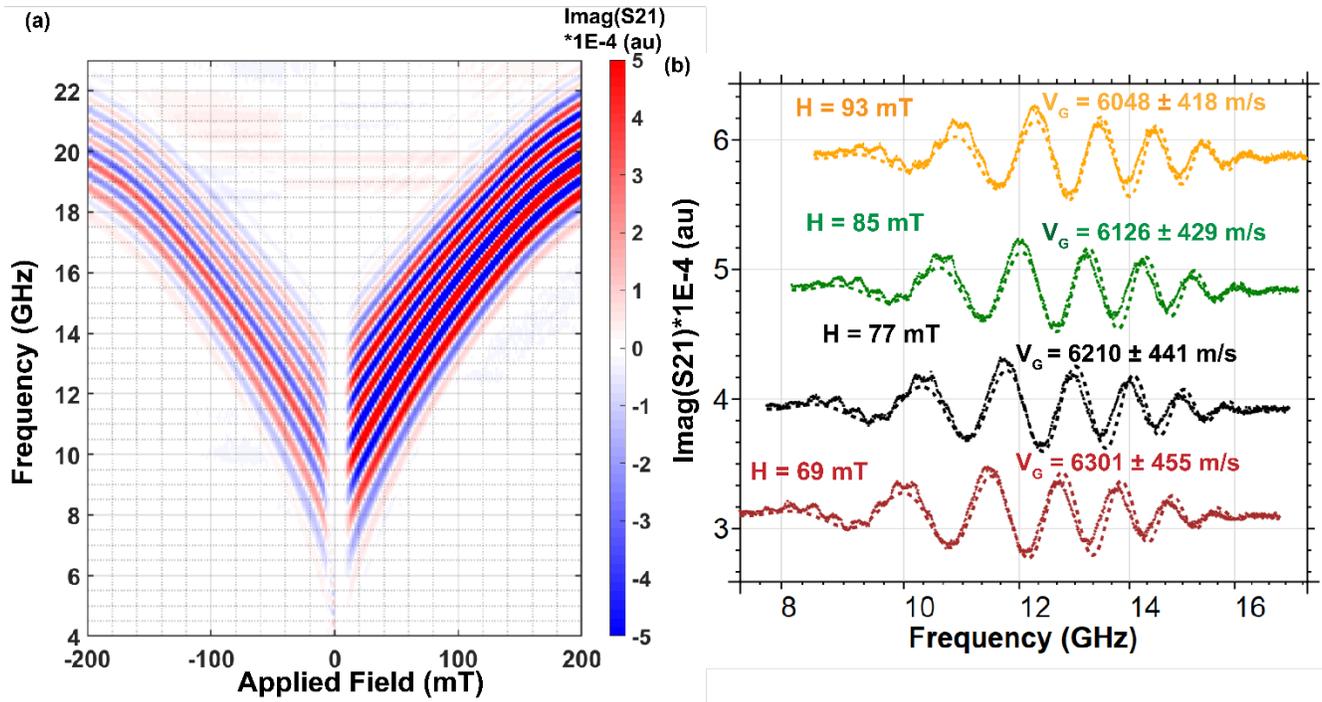

**Figure 2(a):** Frequency versus applied field (FF) map of DESWs for a propagation distance of 6.75 μm; **(b)** Line cuts of the FF map vertically offset for several values of the internal magnetic field H as defined in Eq. 2. The solid lines correspond to experimental measurements, and the dotted lines correspond to the analytical estimates of $S_{21}$ according to Eq. 3 from the fitted dispersion relation. The velocity numbers stand for the mean group velocity and the standard deviation thereof in the 4 to 6 rad/um wavevector interval.

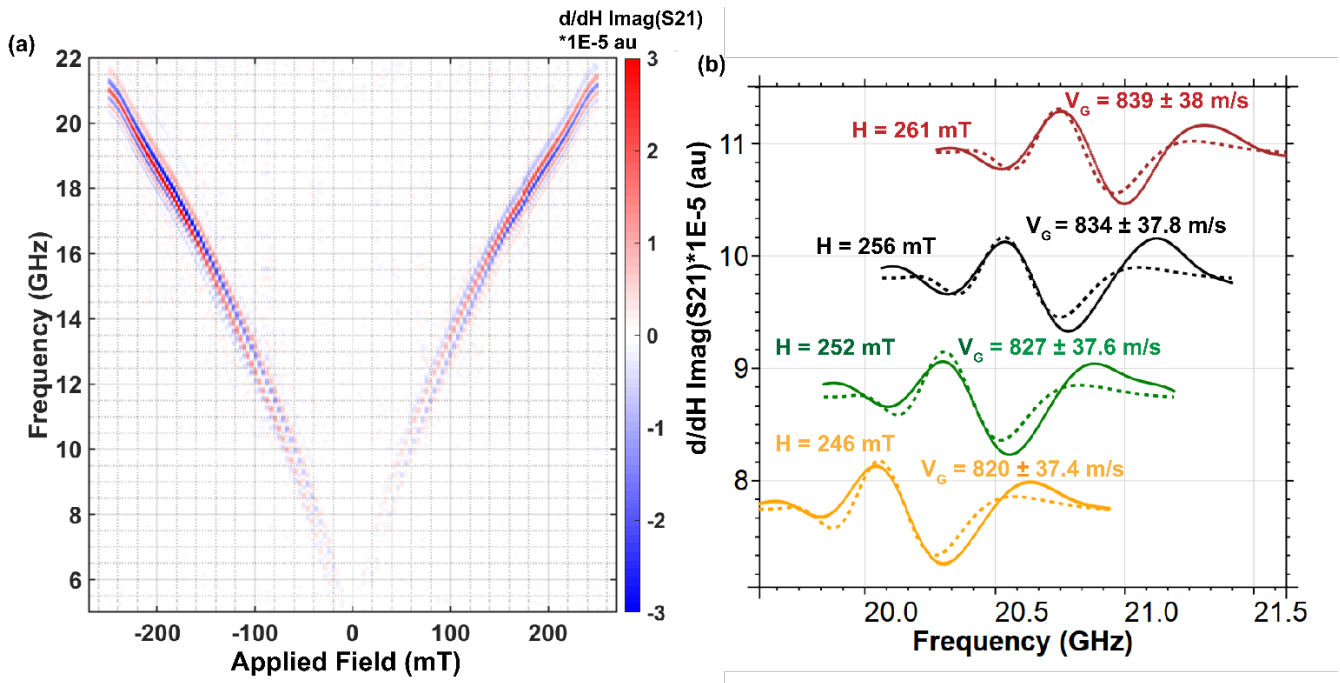

**Figure 3(a):** Frequency versus applied field (FF) map of the field derivative of the BVSWs transmission signal for a propagation distance of 2.25 μm; **(b)** Line cuts of the FF map offset for selected values of the internal magnetic field H as defined in Eq. 2: the solid line corresponds to the experimental measurements, and the dotted line correspond to the analytical estimates of the field derivative of $S_{21}$ according to the field devirative of Eq. 3 from the fitted dispersion relation.

# Backward volume vs Damon-Eshbach:
# a travelling spin wave spectroscopy comparison

U.K. Bhaskar,[1*] G. Talmelli,[2,3] F. Ciubotaru,[3] C. Adelmann,[3] and T. Devolder[1]

**Supplementary information figures**

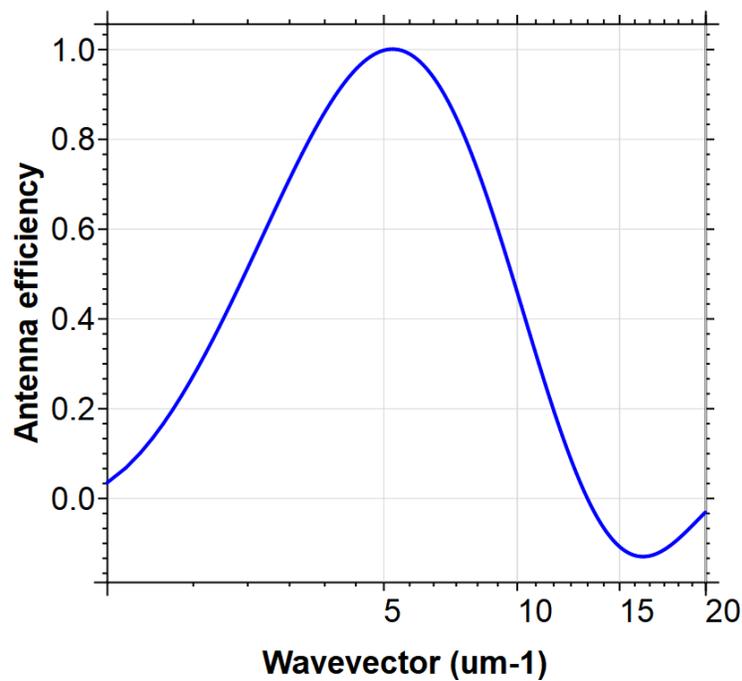

**Supplementary Fig 1** Antenna efficiency for the fabricated width (*w = 250 nm*) and gap-spacing (*gap = 250 nm*) is plotted as a function of wavenumber showing a clear peak around 6 um$^{-1}$.

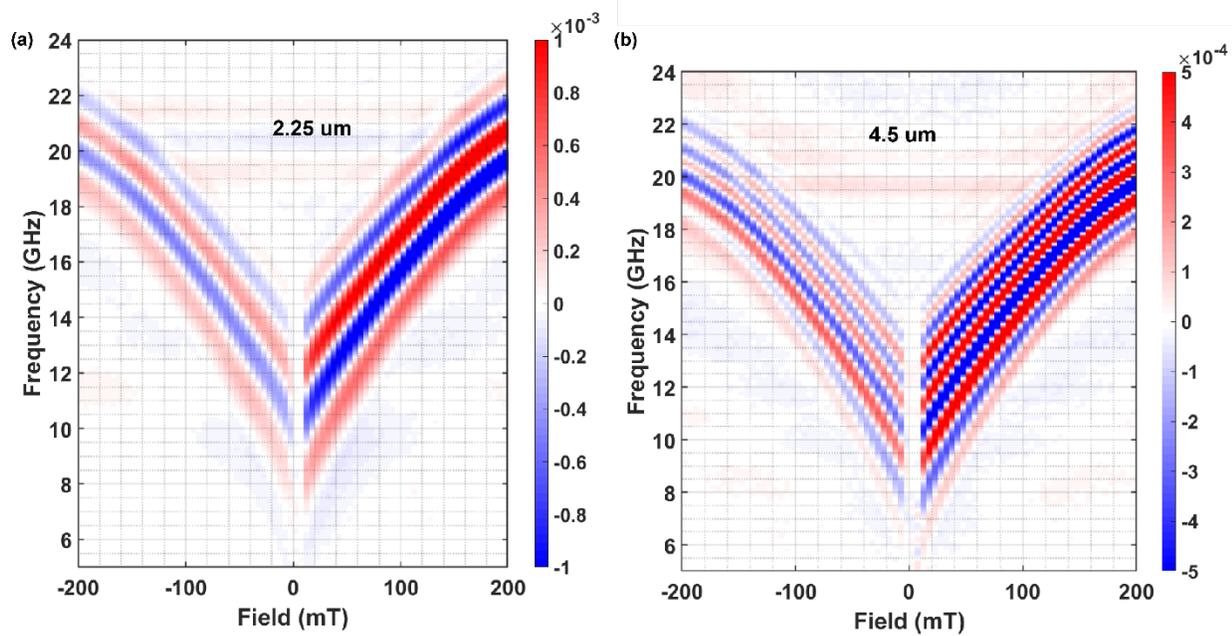

**Supplementary Fig 2** Frequency-field map of DESWs for propogation distances of **(a)** 2.25 µm and **(b)** 4.5 µm respectively.

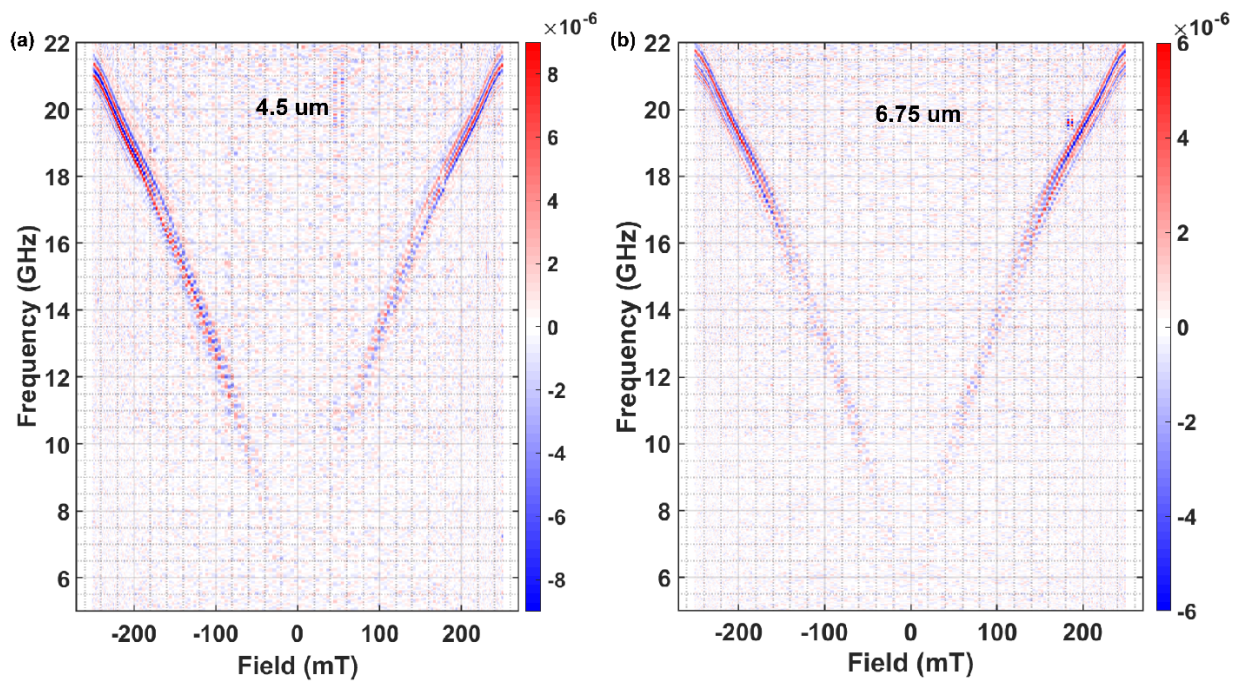

**Supplementary Fig 3** Frequency-field map of BVSWs for propogation distances of **(a)** 4.5 µm and **(b)** 6.75 µm respectively.